\documentclass[11pt]{article}

\usepackage{graphicx}
\usepackage{amsmath,amsfonts,amssymb}
\usepackage{fullpage}

\newcommand\eqnref[1]{(\ref{#1})}
\newcommand\figref[1]{Fig.~\ref{#1}}

\newcommand\sectref[1]{Section~\ref{#1}}

\newcommand{\iu}   {\mathrm{i}}     
\newcommand{\Deltarm}   {\mathrm{\Delta}}

\newcommand{\Omegarm}   {\mathrm{\Omega}}

\newcommand{\bfH}   {\mathbf{H}}
\newcommand{\bfB}   {\mathbf{B}}
\newcommand{\bfE}   {\mathbf{E}}
\newcommand{\bfD}   {\mathbf{D}}
\newcommand{\bfb}   {\mathbf{b}}
\newcommand{\bfd}   {\mathbf{d}}
\newcommand{\bfe}   {\mathbf{e}}
\newcommand{\bfh}   {\mathbf{h}}

\newcommand{\bfk}   {\mathbf{k}}

\newcommand{\bfr}   {\mathbf{r}}
\newcommand{\bfv}   {\mathbf{v}}

\newcommand{\bfw}   {\mathbf{w}}

\newcommand{\epseff}  {\epsilon_{\mathrm{eff}}}
\newcommand{\mueff}   {\mu_{\mathrm{eff}}}
\newcommand{\EPER}   {E_{\mathrm{PER}}}

\begin{document}

\title{Homogenization of Metamaterials by Dual Interpolation of Fields:\\
a Rigorous Treatment of Resonances and Nonlocality}

\author{Igor Tsukerman\\
Department of Electrical and Computer Engineering,\\
The University of Akron, OH 44325-3904, USA\\
igor@uakron.edu
}





\maketitle

%
%

\textbf{Abstract}.
The paper extends and enhances in several ways the recently proposed
homogenization theory of metamaterials [J. Opt. Soc. Am. B 28, 577 (2011)].
The theory is based on a direct analysis of fields in the lattice cells rather than
on an indirect retrieval of material parameters from transmission / reflection data.
The theory is minimalistic, with only two fundamental premises at its core:
(i) the coarse-grained fields satisfy Maxwell's equations and boundary conditions exactly;
and (ii) the material tensor is a linear relationship between the pairs of coarse-grained
fields. There are no heuristic assumptions and no artificial averaging rules.
Nontrivial magnetic behavior, if present, is a logical consequence of the theory.
The method yields not only all 36 standard material parameters, but also additional ones
quantifying spatial dispersion rigorously.
The approximations involved are clearly identified. A tutorial example and an
application to a resonant structure with high-permittivity inclusions are given.

\section{Introduction}\label{sec:Intro}
%
%
The complex electromagnetic behavior of metamaterials -- artificial periodic structures
with features smaller than the vacuum wavelength -- has been extensively investigated
over the last decade. One particularly intriguing phenomenon is
``artificial magnetism'' at high frequencies.
A physical explanation for it is that resonating elements in metamaterial cells
act as elementary magnetic dipoles. However, there is a paradox associated with such magnetism.
In a metamaterial composed of intrinsically nonmagentic components,
the ``microscopic'' (pointwise) magnetic fields $\bfh$ and $\bfb$
are the same\footnote{In the Gaussian system.
In SI, there is the $\mu_0$ factor that is of no principal significance.};
yet somehow their spatial averages --
the respective coarse-grained fields $\bfH$ and $\bfB$ -- differ.
How can two identical quantities give rise to two different averages?

The theory proposed in \cite{Tsukerman-JOSAB11} resolves this paradox. A key observation --
critical from both mathematical and physical viewpoints -- is that, for Maxwell's equations
and standard boundary conditions to be honored, the coarse-grained fields $\bfH$ and $\bfB$
must be obtained from $\bfb$ via \emph{different} interpolation procedures.
Indeed, $\bfH$ has tangential continuity across all interfaces
whereas $\bfB$ has normal continuity.
Specific interpolation procedures producing fields with the required types of continuity
are discussed in \cite{Tsukerman-JOSAB11} and are, for the sake of completeness,
reviewed in \sectref{sec:Proposed-theory-details}.


The theory of \cite{Tsukerman-JOSAB11}, significantly extended and enhanced in the present paper,
has several distinguishing features. First, it is based on a \emph{direct analysis}
of the field in the lattice cell. This is in contrast with S-parameter retrieval procedures
\cite{Smith-PRB02}--\cite{Chen-PRE04} where the effective parameters are inferred
from reflection and transmission coefficients of a metamaterial slab.
Second, the theory is \emph{minimalistic, with only two fundamental premises at its core}:
(i) the coarse-grained fields must satisfy Maxwell's equations and boundary conditions;
and (ii) the material tensor is a linear relationship between the pairs of coarse-grained
fields $(\bfE, \bfH)$ and $(\bfD, \bfB)$. There are no heuristic
assumptions and no artificial averaging rules contrived to arrive at a desired result
such as magnetic permeability $\mu \neq 1$; nontrivial magnetic behavior,
if present, follows logically from the method,
along with other essential characteristics and effects.
Third, not only all 36 standard material parameters, but also additional ones
quantifying spatial dispersion can be found, as explained in
Sect.~\ref{sec:Proposed-theory-details}, \ref{sec:Spatial-dispersion}.

%
%
The effective medium description of metamaterials is, despite very intensive studies,
still far from being settled in a rigorous way. The existing literature on this subject is quite vast,
and the approaches include retrieval via S-parameters, applications and extensions of
classical mixing formulas for small inclusions, analysis of dipole lattices,
special current-driven models, and much more. Reviews and references are available in
\cite{Sarychev-Shalaev-book07}--\cite{Simovski-Tretyakov10}
and a recent summary can be found in the Introduction of \cite{Fietz-PRB10,Tsukerman-JOSAB11}.
Here I summarize only two approaches most relevant to the present paper.

In one common procedure already noted, effective parameters of a metamaterial slab
are ``retrieved'' from transmission/reflection data (i.e. from S-parameters). Being essentially an inverse problem,
this parameter retrieval has some inherent ill-posedness that manifests itself
in the multiplicity of solutions,
due to the ambiguity of branches of the inverse trigonometric functions involved.
Parameters obtained from transmission/reflection from slabs of varying thickness
(let alone shape) are not always consistent. More fundamentally, the retrieval procedure does not
by itself explain why such consistency \emph{should} be expected;
obviously, there must be an underlying reason for it.
That deeper reason is the definition of material parameters
as relations between the (pairs of) coarse-grained fields.

The existing approach most closely related to the methodology of the present
paper is due to the insight of Smith \& Pendry, who suggested
different averaging procedures for different fields in the cell \cite{Smith06}.
Their justification came from the analogy with finite difference schemes on staggered grids.
The present paper, along with \cite{Tsukerman-JOSAB11}, provides a substantially more
rigorous foundation for this physical insight and a much more comprehensive description
of the behavior of the fields in terms of the $6 \times 6$ parameter matrix
and beyond.

The key concepts of the proposed theory are familiar to applied mathematicians
and numerical analysts, but their application to homogenization and metamaterials
is novel. For this reason, the exposition below is split into several parts:
a general description of the key concepts, followed by some technical details,
implementation and examples.


%
\section{Proposed Theory: Key Concepts}\label{sec:Proposed-theory-key-concepts}
The proposed theory follows directly and logically from the definition of material parameters
as linear relations between the coarse-grained fields. Naturally, this requires
these coarse-grained fields to be unambiguously defined
and relations between them established. That is the main theme of this paper.

The motivation for any effective medium theory is that the microscopic fields
vary too rapidly to be easily described and analyzed; one may say that they
have ``too many degrees of freedom''. An obvious idea is to split them up
into coarse-grained (capital letters) and fast
(tilde-letters) components, e.g. $\bfb = \bfB + \bfb^{\sim}$.
For reasons explained in detail below, this splitting should satisfy several conditions:
\begin{enumerate}
  \item
  By definition, the coarse-grained fields must vary much less rapidly than the total fields.
  Nevertheless the coarse-grained fields must approximate, in a certain well defined sense,
  the actual physical fields everywhere in space.
  \item
  The coarse-grained fields must satisfy Maxwell's equations and interface boundary conditions.
  \item
  The fast and coarse components must be semi-decoupled, in the following sense.
  The fast components may depend on the coarse ones, but the coarse fields
  must not depend, or may depend only very weakly, on the fast ones.
  \item
  There exists a linear relationship between the pairs of coarse-grained fields
  $(\bfE, \bfH)$ and $(\bfD, \bfB)$ that
  is independent of the incident waves (at least to a given level of approximation).
\end{enumerate}
The requirements above define not a single method but a \emph{framework}
from which conceptually similar but not fully equivalent homogenization methods could potentially
be obtained by making these requirements more specific.

The rationale for each of the conditions above is as follows. The first one represents,
from the physical perspective, the very essence of homogenization. From a more mathematical viewpoint
of analysis and simulation, one might accept a weaker requirement that the coarse
fields can be described with much fewer degrees of freedom than the total fields;
the coarse fields would not necessarily have to vary less rapidly.
For example, the field in a periodic structure can in many cases be described by
a very limited number of Bloch waves with judiciously chosen wave vectors
(see e.g. \cite{Scheiber-MOR11} for some illuminating examples); this fact
may be effectively used in semi-analytical and numerical methods such as
Generalized Finite Element Method (GFEM) \cite{Babuska97}--\cite{Babuska-Lipton11},
Discontinuous Galerkin methods \cite{Hiptmair11}, and Flexible Local Approximation MEthods (FLAME)
\cite{Tsukerman06,Tsukerman-book07,Tsukerman-PBG08}. Nevertheless
the focus of this paper is on the methods that could give maximum physical insight
via traditional electromagnetic parameters.

\emph{Pointwise} approximation of the miscorscopic fields by the coarse ones
is in general not feasible and not required, as rapid local field
oscillations due to the resonance effects in metamaterial cells
are definitely of interest. Instead, we assume that
the coarse-grained fields are close to the real ones at the cell boundary, where the
fields vary more smoothly. Inside the cell, the coarse-grained fields are defined
by interpolation from the boundary values.


Although the second requirement (coarse-grained fields satisfying Maxwell's equations)
seems to be perfectly natural, many existing methods pay surprisingly little attention to it
and may actually violate it. In particular, coarse fields defined via simple volume averaging
do not satisfy Maxwell's boundary conditions. We shall return to this important point later.

The rationale for the third requirement (weak coupling) is that, if both components were to be
coupled strongly, the two-scale problem would not be any simpler than the original one
involving the total field.

The fourth requirement is for now open-ended and needs to be made more specific.
Let us assume that, for a given microscopic field $(\bfe, \bfd, \bfb)$, the coarse-grained
fields $(\bfE, \bfD, \bfB, \bfH)$ satisfying conditions 1--3 above have been defined in some way.
The central question then is to relate the pairs of coarse-grained fields.
The generalized ``material parameter'' is,
mathematically, a linear map $\mathcal{L}: (\bfE$, $\bfH) \rightarrow (\bfD, \bfB)$
from the functional space of fields $(\bfE$, $\bfH)$ to the functional space of $(\bfD, \bfB)$.
The dimensionality of this map depends on the coarse-grained interpolations chosen.
A high-dimensional linear map could be of use in numerical procedures but does not offer much
physical insight. One critical question then is whether a good \emph{low-dimensional} --
mathematically, a low-rank -- approximation of this linear map can be found.
A pure linear-algebraic answer to this question is well known:
the best approximation of a given rectangular matrix by a matrix of rank $m$
is via the highest $m$ singular values and the respective singular vectors (see e.g.
\cite{Golub96} or \cite{Demmel97} for the mathematical details).
The error of this approximation is equal to the $m+1$st singular value.
One may note a conceptual similarity with the well known Principal Component Analysis (PCA);
see e.g. \cite{Shlens09} for an elementary tutorial.

Although the PCA perspective is instructive, it does not provide a direct connection with
the physical parameters such as $\epsilon$ or $\mu$. Our objective is still to find new bases
in which the material map $\mathcal{L}$ could be ``compressed'' to a low-dimensional form,
but the bases we are seeking are physical rather than formally algebraic.
More specifically, in this physical basis the first few components of the
coarse-grained fields $\bfE, \bfH, \bfD, \bfB$ are to be their mean values,
and the subsequent components -- the increments of these fields.
Such bases and the respective matrix representation of the map $\mathcal{L}$
will be called \emph{canonical}; see \sectref{sec:Proposed-theory-details} for details.


%


%
\section{Proposed Theory: Details}\label{sec:Proposed-theory-details}
\subsection{Equations}
Consider a periodic structure composed of materials that are assumed to be
(i) intrinsically nonmagnetic (which is true at sufficiently high frequencies
\cite{Landau84});
(ii) satisfy a linear local constitutive relation $\bfd = \epsilon \bfe$.
For simplicity, we assume a cubic lattice with cells of size $a$.

Maxwell's equations for the microscopic fields are, in the frequency domain and with the
$\exp(-\iu \omega t)$ phasor convention,
%
$$
 \nabla \times \bfe = \iu \omega c^{-1}\bfb, ~~~
  \nabla \times \bfb = -\iu \omega c^{-1}\bfd
$$
%
%
As in \cite{Tsukerman-JOSAB11}, small letters $\bfb, \bfe, \bfd$, etc.,
will denote the ``microscopic'' -- i.e. true physical -- fields that in general
vary rapidly as a function of coordinates. Capital letters will refer to
smoother fields, to be defined precisely later, that vary on the scale coarser than the lattice
cell size.

The coarse-grained fields $\bfB$, $\bfH$, $\bfE$, $\bfD$ must
be defined in such a way that the boundary conditions be honored.
As noted in \cite{Tsukerman-JOSAB11}, simple cell-averaging does not
satisfy this condition. To understand why, consider, say, the ``microscopic''
magnetic field, for simplicity just a function of one coordinate,
$\bfb(x)$, at a material/air interface $x = 0$. For intrinsically
nonmagnetic media, this field is continuous across the interface, i.e.
$\bfb(x = 0-) = \bfb(x = 0+)$. Since the field fluctuates in the material,
there is no reason for its value at $x = 0$ to be equal to its cell average
over $0 \leq x \leq a$. Thus, if this cell average is used to define
the $\bfB$ field, its normal component at the interface will in virtually
all cases be discontinuous, which is nonphysical. Likewise, if the $\bfH$
field is defined via the cell average, its tangential component will
in general behave in a nonphysical way.

As a rough, ``zero-order,'' approximation, these nonphysical field jumps
across the interface could be neglected. This may indeed be possible in
the absence of resonances or for vanishingly small cell sizes.
However, neglecting strong field fluctuations in other cases
would eliminate the very resonance effects that we are trying to describe.

\begin{figure}
	\centering
    \includegraphics[width=0.6\linewidth]{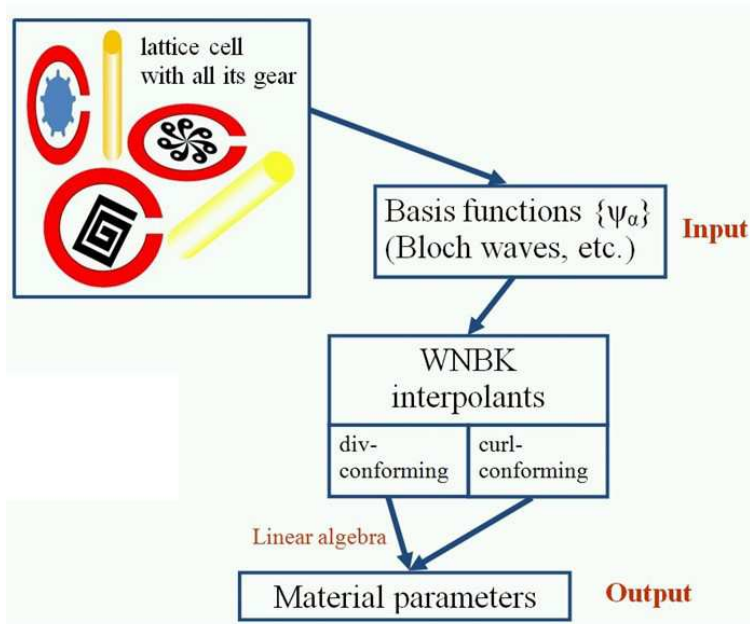}
    \caption{Key parts of the proposed methodology. Two types of interpolation
    are used to obtain the coarse-grained fields with  tangential and normal
    continuity. The electromagnetic field inside the cell, with all its microstructure,
    is approximated with a set of basis modes. Material parameters are linear-algebraic
    relationships between the pairs of coarse-grained fields.}
	\label{fig:Scheme-proposed-methodology}
\end{figure}

To help the reader navigate the proposed procedure without getting bogged down in technical
detail, let us review the general stages of the method (\figref{fig:Scheme-proposed-methodology})
and explain why each of them is necessary. The core of the theory proposed in \cite{Tsukerman-JOSAB11}
-- tangentially- and normally-continuous interpolations,
as well as approximation of the field as a linear combination of basis functions (modes) --
remains intact, but the ``inner workings'' of various components
have been modified, as explained in the subsequent sections.


\subsection{Coarse-grained fields: interpolation and continuity conditions}\label{sec:Vectorial-interpolation}
%
This is the central point of the proposed methodology, where we depart from more traditional
methods of field averaging.\footnote{The interpolation described in this paper is well known
in numerical analysis but, to my knowledge, has not been previously used
for deriving effective parameters.} The coarse-grained $\bfE$ and $\bfH$ fields
are produced from the ``microscopic'' ones, $\bfe$ and $\bfb$, by an interpolation
that respects tangential continuity across all interfaces. The coarse-grained $\bfD$
and $\bfB$ fields are produced from $\bfd$ and $\bfb$ by another interpolation,
one that preserves normal continuity.

\begin{figure}
  \centering
  \includegraphics[width=0.45\linewidth]{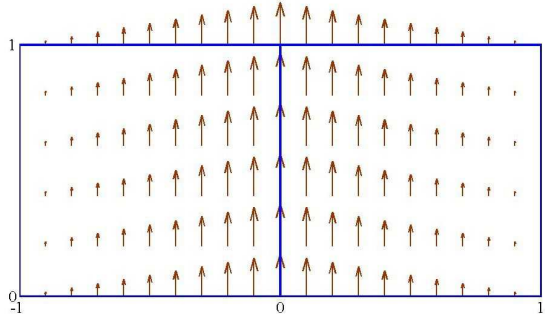}\\
  \caption{(From [Tsukerman, J Opt Soc Am B, \textbf{28}, 2011].) 
  A 2D analog of the vectorial interpolation function $\bfw_\alpha$
  (in this case, associated with the central vertical edge shared by two adjacent cells).
  Tangential continuity of this function is evident from the arrow plot;
  its circulation is equal to one over the central edge and to zero
  over all other edges.}\label{fig:vector-edge-function}
\end{figure}

Tangentially continuous interpolation is effected by
vectorial functions like the one shown in \figref{fig:vector-edge-function},
in a 2D rendition for simplicity. The circulation of this function is equal to one
along one edge (in the figure, the vertical edge shared by two adjacent lattice cells)
and zero along all other edges of the lattice.

To shorten all interpolation-related expressions, \emph{in this section the lattice
cell is normalized to the unit cube} $[0, 1]^3$, i.e. $a = 1$.
Then a formal expression for four $x$-directed functions $\bfw$ is 



%
\begin{equation}\label{eq:curl-conforming-funcs-w}
   \mathbf{w}_{1-4} = 2 \hat{x}
   \{ yz, (1-y)z, y(1-z), (1-y)(1-z) \}
\end{equation}
%

Another eight functions of this kind are obtained by the cyclic permutation of
coordinates in the expression above. For each lattice cell, there are 12 such
interpolating functions altogether (one per edge). Each
function $\mathbf{w}_\alpha$ has unit circulation along edge $\alpha$ ($\alpha = 1,2,\ldots12$)
and zero circulations along all other edges. All of them are vectorial interpolating functions,
bilinear with respect to the spatial coordinates.
(In \cite{Tsukerman-JOSAB11} these functions were defined in the cell scaled to $[-1, 1]^3$
rather than to $[0, 1]^3$, and therefore the algebraic expressions differ.)

%

The coarse-grained $\bfE$ and $\bfH$ fields can then be represented
by interpolation from the edges into the volume of the cell as follows:
%
\begin{equation}\label{eqn:hexahedral-edge-interpolation-EH}
    \bfE ~=~ \sum\nolimits_{\alpha=1}^{12} \, [\bfe]_\alpha \bfw_\alpha,
    ~~
    \bfH ~=~ \sum\nolimits_{\alpha=1}^{12} \, [\bfb]_\alpha \bfw_\alpha
\end{equation}
%
\noindent where $[\bfe]_\alpha = \int_\alpha {\bfe \cdot \mathbf{dl}}$
is the circulation of the (microscopic) $\bfe$-field along edge $\alpha$;
similarly for $[\bfb]_\alpha$.
In the calculation of the circulations, integration along the edge
is always assumed to be in the positive direction of the respective coordinate axis.

\begin{figure}
  \centering
  \includegraphics[width=0.45\linewidth]{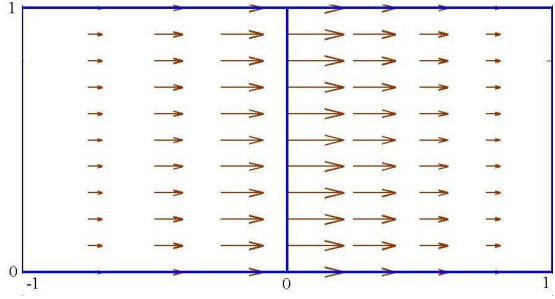}\\
  \caption{(From [Tsukerman, J Opt Soc Am B, \textbf{28}, 2011].) 
  A 2D analog of the vectorial interpolation function $\bfv_\beta$
  (in this case, associated with the central vertical edge).
  Normal continuity of this function is evident from the arrow plot;
  its flux is equal to one over the central edge and zero
  over all other edges.}\label{fig:vector-face-function}
\end{figure}

Now consider the second kind of interpolation that preserves the normal continuity
and produces the $\bfD, \bfB$ fields from $\bfd$ and $\bfb$. A typical interpolating function
(2D rendition again for simplicity) is shown in \figref{fig:vector-face-function}.
The flux of this function through a face shared by two adjacent cells is equal to one;
the flux through all other faces is zero. Two such functions in the $x$-direction are
(as before, for the cell size normalized to unity)
$$
    \bfv_{1-2} = \hat{x} \{ x, 1-x \}
$$
%
and another four functions $\bfv_{3-6}$, in the $y$- and $z$-directions, are
expressed similarly. These six functions can be used to define the coarse-grained $\bfD$ and $\bfB$ fields
by interpolation from the six faces into the volume of the unit cell:
\begin{equation}\label{eqn:hexahedral-edge-interpolation-DB}
    \bfD \,=\, \sum_{\beta=1}^{6} [[d]]_\beta \mathbf{v}_\beta,
    ~~~
    \bfB \,=\, \sum_{\beta=1}^{6} [[b]]_\beta \mathbf{v}_\beta,
\end{equation}
where $[[d]]_\beta = \int_\beta{\mathbf{d}\cdot \mathbf{dS}}$
is the flux of $\bfd$ through face $\beta$ ($\beta = 1,2,\ldots,6$);
similar for the $\bfb$ field. In the calculation of fluxes, it is convenient
to take the normal to any face in the positive direction of the respective coordinate axis
(rather than in the outward direction).

The coarse-grained $\bfE$ and $\bfH$ fields so defined have 12 degrees of freedom in any given lattice cell.
From the mathematical perspective, these fields lie in the 12-dimensional functional space
spanned by functions $\bfw_{\alpha}$; we shall keep
the notation $W_\mathrm{curl}$ of \cite{Tsukerman-JOSAB11} for this space:
the `W' honors Whitney \cite{Whitney57}
(see \cite{Tsukerman-JOSAB11} for details and further references)
and `curl' indicates fields whose curl is a regular function
rather than a general distribution. This implies, in physical terms,
the absence of equivalent surface currents and the tangential continuity of the fields involved.

Similarly, $\bfD$ and $\bfB$ within any lattice cell lie in the six-dimensional
functional space $W_\mathrm{div}$ spanned by functions $\bfv_{\beta}$.
Importantly, it can be shown that the div- and curl-spaces are compatible in the following sense:
\begin{equation}\label{eqn:curl-W-in-Wdiv}
    \nabla \times W_\mathrm{curl} \in W_\mathrm{div}
\end{equation}
That is, the curl of any function from  $W_\mathrm{curl}$ (i.e. the curl of
any coarse-grained field $\bfE$ or $\bfH$ defined by \eqnref{eqn:hexahedral-edge-interpolation-EH})
lies in $W_\mathrm{div}$. Because of this compatibility of interpolations,
the coarse-grained fields, as proved in \cite{Tsukerman-JOSAB11}, satisfy
Maxwell's equations exactly:
\begin{equation}\label{eqn:curl-E-eq-dBdt-coarse}
    \nabla \times \bfE  \,=\, \iu \omega c^{-1} \bfB; ~~
    \nabla \times \bfH \,=\, -\iu \omega c^{-1} \bfD
\end{equation}
By construction, they also satisfy the proper continuity conditions at all interfaces.
Thus the coarse-grained fields are completely physical.

\begin{figure}
	\centering
    \includegraphics[width=0.5\linewidth]{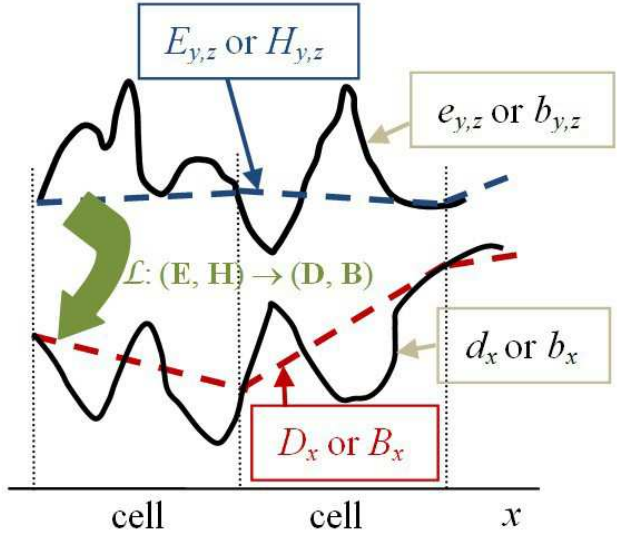}
    \caption{A linear relation between the coarse-grained fields. $\bfE, \bfH$
    are interpolated to ensure tangential continuity; $\bfD, \bfB$ are normally continuous.
    1D rendition for simplicity. The linear relation $\mathcal{L}$ between the pairs of fields
    is in general multidimensional but in ``homogenizable'' cases has a dominant $6 \times 6$
    (or smaller) block.}
	\label{fig:coarse-grained-fields-interpolation}
\end{figure}

To recap the ideas, \figref{fig:coarse-grained-fields-interpolation}
schematically illustrates the interpolation procedures and the linear-algebraic
part of the proposed method. The simplified 1D rendition shows only the $x$ axis,
the tangential ($y, z$) components of $\bfE, \bfH$ and the normal ($x$)
component of $\bfD, \bfB$. Other components (not shown) may be discontinuous
at cell boundaries and at material interfaces.

The linear relation $\mathcal{L}$ between the pairs of coarse-grained fields is in general multidimensional,
commensurate with the dimensions of the interpolation spaces chosen.
In ``homogenizable'' cases, $\mathcal{L}$ has a dominant $6 \times 6$ (or smaller)
matrix block in a suitable ``canonical'' basis; this block relates the
field averages of $(\bfD, \bfB)$ to the field averages of $(\bfE, \bfH)$ .
The remainder of the matrix relates the field averages to field \emph{variations}
over the cell and can be viewed as a manifestation of nonlocality
(see \sectref{sec:Spatial-dispersion} for further discussion).

%


%
\subsection{Approximating Functions}\label{sec:Approx-functions}
%
The actual microscopic fields in the material are in general not known and
need to be approximated by introducing suitable basis functions \cite{Tsukerman-JOSAB11}.
For maximum generality, the choice of the basis is left open for now.
Examples inclide Bloch waves in the periodic case, plane waves, cylindrical or
spherical harmonics, etc.
Although Bloch modes are very useful and
physically meaningful, their use does have disadvantages and limitations. They are rigorously
applicable only to periodic structures, which is strictly speaking not the case if the
effective properties of the metamaterial need to vary -- e.g. in cloaking applications.
Evanescent field components are approximated by traveling Bloch waves very poorly,
but these components are essential in imaging and other applications.
In addition, Bloch modes are expensive to compute. An alternative set of basis functions that
does not require eigenmodes is introduced below.

For a non-vanishing cell size, the field has an infinite number of degrees of freedom;
thus, as a matter of principle, its representation in a finite basis is approximate. Nevertheless
the relevant behavior of the fields can be captured very well by relatively small bases.
By expanding the bases, one can increase the accuracy
of the representation of the fields, at the expense of higher complexity.
A small number of degrees of freedom corresponds to local material parameters,
a moderately higher number -- to nonlocal ones, and a large number falls under the umbrella
of numerical simulation. This \emph{unifies local, nonlocal and numerical
treatment} of electromagnetic characteristics of metamaterials.

Note that tangential components of the electric or, alternatively, magnetic field
on the cell boundary define the field inside the cell uniquely, except for the special cases
of interior resonances. To specify the basis, it is thus sufficient to consider the boundary
values in lieu of the values inside of the cell.

With all the above in mind, a rather natural set of approximating functions
contains EM fields with the boundary values of $\bfE$ or $\bfH$
equal to those of the curl-conforming functions \eqnref{eq:curl-conforming-funcs-w}.
In this setup, the approximating functions
for the $\bfe$ and $\bfb$ fields \emph{on the cell boundary} are the same
as the interpolating functions for the coarse-grained fields $\bfE, \bfH$.
(The normally-continuous functions are still used for $\bfD, \bfB$.)

Since the chosen interpolation functions are low-order polynomials, the accuracy of this approximation
will depend on the smoothness of the field on the cell boundary and in general can be expected
to depend on the size of the ``buffer zone'' between the inclusions in the cell and its boundary.
However, effective parameters are essentially a zero-order approximation of the fields:
a relationship between averaged fields that vary slowly within the cell.
From this perspective, the low-order polynomial approximation on the boundary
is unlikely to be a limiting factor in most practical cases.

The reader well versed in numerical analysis will immediately recognize that
this type of interpolation is borrowed directly from edge elements in the
finite element method (FEM), \cite{Bossavit98}.
However, our goal is to develop an \emph{analytical}
rather than numerical procedure. It is curious, though not accidental,
that the ``numerical'' tools of FEM prove to be quite suitable for analytical purposes
as well.

It remains to decide whether the boundary values of the $\bfe$ field, $\bfb$ field or both
will be taken as a basis for defining the approximating set. If one of the fields is chosen
(either $\bfe$ or $\bfb$, arbitrarily), the number of approximating functions is equal to
the number of curl-conforming functions \eqnref{eq:curl-conforming-funcs-w},
i.e. to 12. If both fields are chosen,
there are 24 functions in the approximating set. In this latter case, some redundancy is apparent
because the electric and magnetic fields are not independent. Nevertheless, for reasons of symmetry
and elegance, it is convenient to use the 24-function set and deal with the redundancy later.

%
\subsection{The Canonical Relation and Material Parameters}\label{sec:Cell-response}
%
By construction of the coarse-grained interpolants, the $\bfE$ and $\bfH$ fields
taken together lie in a 24-dimensional linear space (12 edge-based degrees of freedom for each field),
while the $\bfD$ and $\bfB$ fields together lie in a 12-dimensional space (six face-based
degrees of freedom for each field). Therefore a linear relationship between these fields
is, technically, a map from the space 
of complex 24-component vectors to the space of complex 12-component vectors; 
for a given pair of bases in these spaces, this map is described by
a $24 \times 12$ complex matrix.

This description is, however, cumbersome and redundant.
First, as already noted, the electric and magnetic fields are not independent,
so the information contained in the 12 d.o.f. of the $\bfE$ field is similar
to that of the 12 d.o.f. of the $\bfH$ field. Secondly, and more importantly,
the fields can be decomposed into constant components (i.e. zero-order terms
with respect to the cell size $a$) plus progressively less significant contributions
corresponding to the increasing orders of $a$. This is analogous to the Taylor
expansion of the field components, although a more accurate analogy would be with
the Newton divided difference formula (see e.g. \cite{deBoor05}).

\emph{Remark 1}. The Taylor expansion is fully valid for the coarse-grained fields
but not for the microscopic ones,
as the latter are not smooth due to the jumps across various material interfaces.
The Newton formula applies in all cases.


A natural, ``canonical,'' basis then presents itself.
Consider the average tangential components $e_{z1}, e_{z2}, e_{z3}, e_{z4}$
of the electric field along the four edges of the lattice cell aligned in the $z$ direction.
Two vector components in the canonical basis are the mean value plus the increment (difference):
\begin{equation}\label{eqn:edge-to-hierarchical-basis-ez-mean}
   e_{z0} = (e_{z1} + e_{z2} + e_{z3} + e_{z4})/4
\end{equation}
\begin{equation}\label{eqn:edge-to-hierarchical-basis-Deltax-ez}
   \Deltarm_x e_z  \equiv \left( \frac{e_{z2} + e_{z3}}{2} - \frac{e_{z1} + e_{z4}}{2} \right) / (ka)
\end{equation}
%
%
(The normalization by $ka$ keeps the field increments at the same order of magnitude
as the fields themselves if the wavenumber and/or the cell size tend to zero.)
Another two vector components are obtained by replacing $z$ with $x$ and $y$ in
\eqnref{eqn:edge-to-hierarchical-basis-ez-mean} and yet another two by the cyclic permutation
of coordinates in \eqnref{eqn:edge-to-hierarchical-basis-Deltax-ez}.
The total number of vector components related to the electric field is then six.
Similarly, there are six components related to $h$.

The expressions above constitute a transformation from the
edge-based values of the fields to their more physical representation in terms of the mean values
and increments. Note that some redundancy has been implicitly eliminated at this stage:
the 24 degrees of freedom (12 edge values for each field) have been reduced to
the total of 12. In principle, the dimension of 24 could be preserved by including
higher-order differences in the canonical basis, but this would only perpetuate
the redundancy by retaining unnecessary terms in the field expansion.


\emph{Remark 2}. The differences $\Deltarm$ correspond to the respective partial derivatives
of the field, but only in a smoothed, average sense (see Remark 1).

A similar basis change can be considered for the $\bfD$ and $\bfB$ fields:
\begin{equation}\label{eqn:face-to-hierarchical-basis}
    d_{z0} = (d_{z1} + d_{z2})/2; ~~
    \Deltarm_z d_z = (d_{z1} - d_{z2})/(ka);
   ~~ \mathrm{etc.}
\end{equation}
where $d_{z1}$ and $d_{z2}$ are the average $z$-components of the $\bfd$ field
on the two faces normal to the $z$ axis. Similar expressions apply in the other two
coordinate directions, bringing the total number of transformation formulas to six
for each of the two fields $\bfd$ and $\bfb$.

The linear algebra part of the procedure is conceptually similar to the one
in \cite{Tsukerman-JOSAB11}, but with several enhancements leading to a
rigorous and quantifiable definition of spatial dispersion.
Let the electromagnetic field be approximated as a linear combination of some
basis waves (modes) $\psi_\alpha$:
$$
    \Psi^{eh} \,=\, \sum\nolimits_\alpha c_\alpha \psi_\alpha^{eh}; ~~~~
    \Psi^{db} \,=\, \sum\nolimits_\alpha c_\alpha \psi_\alpha^{db}
$$
In the most general case, $\Psi$ and all $\psi_\alpha$ are six-component vector
comprising both microscopic fields; e.g.
$\Psi^{eh} \equiv \{ \Psi^e, \Psi^h \}$, etc.
For periodic structures the most natural choice for the basis waves $\psi_\alpha$ is
Bloch modes as stated in \cite{Tsukerman-JOSAB11}; however, in this paper,
as already noted, we are also interested in defining the approximating functions
via their boundary values equal to Whitney interpolants.

Each approximating function $\psi_\alpha$ can be expressed in the edge or face
Whitney basis on the cell boundary. This representation can then be converted,
using the transformations above,
to the canonical form. More precisely, for each approximating wave one first computes
the 12 edge circulations of each of the $\bfe$ and $\bfb$ fields; the resulting 24-vector
is then transformed to the canonical basis using
\eqnref{eqn:edge-to-hierarchical-basis-ez-mean}, \eqnref{eqn:edge-to-hierarchical-basis-Deltax-ez}.
Likewise, for each approximating wave one first computes the six face fluxes of each
of the $\bfd$ and $\bfb$ fields; the resulting 12-vector is then transformed into the canonical basis
according to \eqnref{eqn:face-to-hierarchical-basis}.

We then seek a linear relation
\begin{equation}\label{eqn:Psi-DB-eq-zeta-Psi-EH}
      \eta \Psi^{EH} ~=~ \Psi^{DB}
\end{equation}
where each column of the matrices $\Psi^{DB}$ and $\Psi^{EH}$ contains
the respective basis function represented in the canonical basis.
Unlike in the approach taken in \cite{Tsukerman-JOSAB11},
the quantities in this relationship are not position-dependent.
The dimensions of the matrices $\eta$, $\Psi^{EH}$ and $\Psi^{DB}$
are $12 \times 24$, $24 \times N_m$, and $12 \times N_m$,
where $N_m$ is the number of approximating modes.

If $\Psi^{EH}$ is a square matrix,
one obtains the constitutive matrix $\eta$ by straightforward matrix inversion;
otherwise $\eta$ is defined as the pseudoinverse \cite{Golub96}:
\begin{equation}\label{eqn:zeta-eq-Psi-DB-Psi-EH-plus}
      \eta ~=~ \Psi^{DB} \, (\Psi^{EH})^+
\end{equation}
The structure and physical meaning of this matrix are discussed in the following section
in connection with spatial dispersion.

%
\section{``Spatial Dispersion:'' the Wheat and the Chaff}\label{sec:Spatial-dispersion}
%
The goal of this section is to recap the notions of nonlocality / spatial dispersion
for natural materials and to see to what extent these notions apply to metamaterials.
For brevity of notation, this section deals only with the $(\bfe, \bfd)$ pair of fields,
even though similar considerations apply to constitutive relationship between all fields.

Consider first an \emph{infinite homogeneous} medium with a nonlocal spatial response
(in the frequency domain) of the form
$$
    \bfd(\bfr) ~=~ \int_{\bfr'} \epsilon(\bfr - \bfr') \, \bfe(\bfr') d \bfr'
$$
Dependence of all variables on frequency $\omega$ is suppressed to shorten the notation.
The integral is taken over the whole space.

The Fourier transform simplifies this relationship drastically by turning
the convolution into multiplication:
$$
    \bfd(\bfk) ~=~ \epsilon(\bfk) \, e(\bfk)
$$

However, if the medium is \emph{not} homogeneous in the whole space, the translational
invariance is broken and the nonlocal convolutional response becomes a function
of two position vectors independently, rather than just of their difference
\cite{Agranovich84,Markel-private10}:
$$
    \bfd(\bfr) ~=~ \int \epsilon(\bfr, \bfr') \, \bfe(\bfr') d \bfr'
$$
This expression is no longer simplified by the Fourier transform,
and the permittivity can no longer be expressed
as a function of a single $\bfk$-vector in Fourier space.

Since metamaterials are obviously always finite in spatial extent, spatial dispersion
in them, strictly speaking, cannot be described by the dependence
of effective material parameters (however defined) on a single $\bfk$-vector.
If this limitation is ignored, the practical result is likely to be
``smearing'' of material characteristics at the material-air interfaces
and violation of the Maxwell boundary conditions.

Still, in many publications material parameters are derived for a single wave
-- either a Bloch wave or a wave generated by some special auxiliary sources
\cite{Silveirinha07,Fietz-PRB10} --
and then dependence of these parameters on the wavevector $\bfk$ is investigated.
This approach raises further questions, in addition to the lack of translational invariance
noted above.

First, fundamentally, electromagnetic parameters cannot be defined only
from waves propagating in the bulk \cite{Tsukerman-JOSAB11}.
This is so because a gauging transformation may change the relationships
between the fields while leaving Maxwell's equations unchanged \cite{Tsukerman-JOSAB11,Vinogradov02}.
As a simple example, given the microscopic physical fields $\bfe$ and $\bfb$
and the auxiliary fields $\bfd$ and $\bfh$, one observes that Maxwell's equations
are invariant with respect to the transformation
$\bfh' = \bfh/2$, $\bfd' = \bfd/2$, $\mu' = 2 \mu$, $\epsilon' = \epsilon/2$,
even though the material parameters have changed by a factor of two.
It is only through the boundary conditions on the material-air (or material / dielectric)
interfaces that the $\bfd$ and $\bfh$ fields are gauged uniquely.

Second, a natural definition of material parameters for a single Bloch wave propagating in
the bulk of an intrinsically nonmagnetic metamaterial yields only a trivial result
for the effective magnetic permeability, and hence artificial magnetism cannot be explained.
To elaborate, consider an $x$-polarized Bloch wave traveling in the $z$ direction
(e.g. \cite{Tsukerman08,Tsukerman-book07}):
\begin{equation}\label{eqn:E-Bloch}
   \bfe_B(z) ~=~ \EPER(z) \exp(\iu K_B z) \hat{x}
\end{equation}
\begin{equation}\label{eqn:H-Bloch}
   \bfb_B(z) = \hat{y}(\iu \omega)^{-1}
   (\EPER'(z) + \iu K_B \EPER(z)) \exp(\iu K_B z)
\end{equation}
Here ``PER'' indicates a factor periodic over the cell; $K_B$ is the Bloch wavenumber.
For this Bloch wave to mimic a plane wave in an equivalent effective medium,
one has to have the dispersion relation
\begin{equation}\label{eqn:Bloch-dispersion-relation}
   \omega^2 \mueff \epseff ~=~ K_B^2
\end{equation}
and the wave impedance
\begin{equation}\label{eqn:Bloch-impedance}
   (\mueff / \epseff)^{\frac12} ~=~ \langle e_{B} \rangle / \langle b_{B} \rangle
   ~=~ \omega / K_B
\end{equation}
where the angle brackets denote an averaging procedure that eliminates the periodic
function $\EPER'$ and leaves only the dc component in $\EPER$.
From the two conditions above, it follows immediately that $\mueff = 1$.

Third, even if all the complications above are somehow circumvented, a single Bloch or plane wave still
does not carry enough information identifying all 36 parameters in the $6 \times 6$
material parameter matrix. For example, a plane wave does not ``probe'' the longitudinal
field components at all.
Progress is made in \cite{Fietz-PRB10} by considering a simultaneous collection of test waves
that allow one to set up a system of equations for all parameters. These test waves,
however, are generated by mathematical sources that do not seem to be physically
realizable. Other objections against the models with such sources have also been raised
\cite{Markel-Current-driven-model-10}.

All of this is not to say that ``spatial dispersion'' is an invalid notion
for metamaterials. To the contrary, the theory proposed in this paper does
lead to its precise and mathematically quantifiable definition.
The message here is that ``spatial dispersion'' is a loaded expression
that should be used with extreme care and with accurate definitions.
Here is one example of a pitfall.
Spatial dispersion is usually defined as dependence of material parameters
on the wave vector (e.g. $\epsilon = \epsilon(\bfk)$). Logically, this requires
that such functional dependence be established first, and only then any conclusions about
the level of spatial dispersion can be made. It is then tempting to consider
a single wave (be that a Bloch wave or in some cases a plane wave)
with a vector $\bfk$, find the material parameters for that wave and investigate their dependence on $\bfk$.
However, as we have seen, material parameters cannot be consistently defined
for just a single wave.

\begin{figure}
  \centering
  \includegraphics[width=0.4\linewidth]{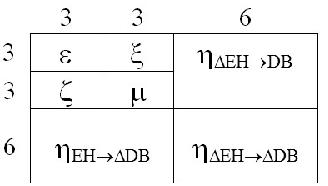}\\
  \caption{The block structure of the extended material parameter matrix, with the numbers of rows/columns in hte blocks shown.}\label{fig:parameter-matrix-blocks}
\end{figure}

The homogenization theory described in the previous sections allows us
to define spatial dispersion in a rigorous and quantifiable way. Indeed,
in the canonical basis the extended material parameter matrix $\eta$
\eqnref{eqn:zeta-eq-Psi-DB-Psi-EH-plus} has the block structure
shown in \figref{fig:parameter-matrix-blocks}.

The leading $6 \times 6$ block contains the standard electromagnetic parameters:
the $\epsilon$ and $\mu$ tensors as well as magnetoelectric coupling
parameters $\xi$ and $\zeta$ (also in general tensorial). The remaining blocks
are novel; they quantify spatial dispersion. In particular,
the $\eta_{\Deltarm \mathrm{EH} \rightarrow \mathrm{DB}}$ block,
as the notation suggests, characterizes the dependence of fields $\bfD$ and $\bfB$
on the \emph{variations} of $\bfE$ and $\bfH$ within the cell rather than on their mean values.
%
\section{Application Examples}\label{sec:Application-examples}
%
\subsection{Tutorial: Homogeneous Cell Revisited}\label{sec:Example-homogeneous-cell}
%
This case, the simplest but still nontrivial, doubles as a tutorial example
where construction of all approximating functions and matrices can be demonstrated
in closed form. In \cite{Tsukerman-JOSAB11,Pors11}, it was shown that
the exact result $\epseff = \epsilon$, $\mueff = 1$ can be obtained
for a homogeneous cell with a permittivity $\epsilon$ and permeability $\mu = 1$
if plane waves are used as the basis modes. However, one cannot expect
such a perfect result in general, due to approximations involved in any
homogenization procedure. It is instructive to see how the proposed methodology
works for the homogeneous cell if Whitney-like values for the basis waves
on the cell boundary are used instead of plane/Bloch waves.

To make the demonstration most transparent, consider the 2D case,
$H$-mode ($p$-mode), governed by the wave equation
$$
    \nabla^2 h \,+\, k^2 h ~=~ 0
$$
where in a homogeneous cell $k$ = const. In this example, as in the section
on interpolating functions, it is convenient to normalize the cell size $a$
to unity, to shorten the notation. Then in the square lattice cell $\Omegarm = [0, 1]^2$,
the four approximating functions $\psi^{(h)}_{1-4}$ for the $h$ field
are chosen to satisfy the wave equation
and Whitney-like conditions on the cell boundary $\partial \Omegarm$
$$
    \psi^{(h)}_1(\partial \Omegarm) = xy; ~~ 
    \psi^{(h)}_2(\partial \Omegarm) = (1-x)y; 
$$
\begin{equation}\label{eqn:bilin-bc-dOmega}
    \psi^{(h)}_3(\partial \Omegarm) = x(1-y); ~~ 
    \psi^{(h)}_4(\partial \Omegarm) = (1-x)(1-y)  
\end{equation}
Inside the cell, these functions can be found using a variety of semi-analytical or numerical methods.
For example, separation of variables leads to the following expression for $\psi^{(h)}_1$
($\psi^{(h)}_{2-4}$ being completely similar):
%
%
$$
    \psi^{(h)}_1 \,=\, xy \,-\,
    k^2 \sum\nolimits_{n_{x,y} = -\infty}^{\infty}
    c_{n_x,n_y} \sin(n_x \kappa x) \sin(n_y \kappa y),
$$
where $\kappa = \pi/a$ in general and $\kappa = \pi$ for the unit cell;
%
$c_{n_x,n_y} = 4 f_{n_x,n_y}/(k^2 - \kappa^2 (n_x^2 n_y^2))$,
$ f_{n_x,n_y} = (\sin \pi n_x - \pi n_x \cos \pi n_x) (\sin \pi n_y -
\pi n_y \cos \pi n_y )/(\pi^4 n_x^2 n_y^2) $.

With the approximating functions $\psi^{(h)}_{1-4}$ for the $h$ field available,
the respective basis waves $\psi^{(e)}_{1-4}$ for the $e$ field
are immediately found by differentiation, from Maxwell's equations.

To find the effective parameter matrix $\eta$ from \eqnref{eqn:Psi-DB-eq-zeta-Psi-EH},
we need to populate the matrices $\Psi^{EH}$ and $\Psi^{DB}$.
The first one contains the edge-averaged $e$ and $h$ fields of the basis waves.
For the 2D case, $p$-mode, the latter degenerate simply to the values of $h$
at the nodes of the lattice cell.
(For the $s$-mode, the circulations of the $e$ field degenerate to the nodal values of the field.)

Matrix $\Psi^{DB}$ contains the face-averaged $\bfd$ and $\bfb$ fields.
For the 2D case, $p$-mode, ``face-averaging'' of the $\bfd$ field turns into edge-averaging
of its component normal to the respective edge, and ``face-averaging'' of the $\bfb$ field turns into
averaging over the whole lattice cell.
(For the $s$-mode, the manipulation of $\bfd$ and $\bfb$ in this procedure would be reversed.)

Let us illustrate this calculation for the first basis mode $\psi^{(h)}_1, \psi^{(e)}_1$.
The four edge-averaged values of the $\bfe$ field are
%
%

$$
    e_{x1} = \int_{x=0}^1 \psi^{(e)}_{1x} (x, 0) dx; ~~~
    e_{x3} = \int_{x=0}^1 \psi^{(e)}_{1x} (x, 1) dx;
$$
$$
    e_{y2} = \int_{y=0}^1 \psi^{(e)}_{1y} (1, y) dy; ~~~
    e_{y4} = \int_{y=0}^1 \psi^{(e)}_{1y} (0, y) dy
$$
The edges are numbered counterclockwise, starting from the edge $[0, 1]$ on the $x$ axis.
The first column of $\Psi^{EH}$ corresponds to the first basis mode,
and the first two entries of that column are
$$
    \Psi^{EH}_{11} \equiv (e_{x1} + e_{x3})/2; ~~
    \Psi^{EH}_{21} \equiv (e_{y2} + e_{y4})/2
$$
The first entry is the mean value of the $x$-components of the electric field along
edges 1 and 3 (the ``horizontal'' edges) of the cell; the second entry is the mean
value of the $y$-components over the ``vertical'' edges 2 and 4.

The third entry of this first column is just the average $h$ field at the nodes
of the cell:
$$
    \Psi^{EH}_{31} ~=~ (h_1 + h_2 + h_3 + h_4)/4
$$
where $h_1 = \psi^{(h)}_1(0,0)$, $h_2 = \psi^{(h)}_1(1,0)$, $h_3 = \psi^{(h)}_1(1,1)$,
$h_4 = \psi^{(h)}_1(0,1)$.

The fourth and final entry in the first column of $\Psi^{EH}$ is
$$(k a)^2 \Psi^{EH}_{41} = h_1 + h_3 - h_2 - h_4 = 0+1-0-0 = 1$$

Computing the first column of the EH matrix as defined above (with a
separation-of-variables expansion of the first basis mode), one obtains
$(0.5 \iu (\epsilon k a)^{-1}, -0.5 \iu (\epsilon k a)^{-1}, 0.25, 1/(ka)^2)^T$,
with $a = 1$ for the unit cell.
The remaining three columns of this matrix correspond to the other three basis waves
and differ from the first one only through cyclic permutation and some sign changes.

Consider now the first column of the DB matrix $\Psi^{DB}$. This column corresponds
to the first basis mode, and other columns are completely analogous.
First, we need the fluxes of the basis wave through the four edges of the cell.
For the bottom edge (edge \#1)
$$
    d_{y1} = a^{-1} \, \epsilon \int_{x=0}^a \psi^{(e)}_{1y} (x, 0) dx
    = 1/(\iu \omega a) (h_2 - h_1)
$$
(with $a = 1$ if the cell is scaled to unit size);
$d_{x2}, d_{x4}$, and $d_{y3}$ are expressed analogously.
%
For the Whitney-type basis functions with bilinear boundary conditions \eqnref{eqn:bilin-bc-dOmega},
these expressions are in fact quite simple. For the first basis wave,
$h_3 = 1$ and $h_1 = h_2 = h_4 = 0$.

The first entry of the first column is the mean value of the $x$-component of the $\bfd$ field along
edges 2 and 4 (the ``vertical'' edges) of the cell, and the second entry
is the mean value of the $y$ component along the horizontal edges:
$$
    \Psi^{DB}_{11} \equiv \langle d_x \rangle \equiv (d_{x2} + d_{x4})/2
$$
$$
    \Psi^{DB}_{21} \equiv \langle d_y \rangle \equiv (d_{y1} + d_{y3})/2
$$
The third entry is the average $B$ field over the cell:
$$
    \Psi^{DB}_{31} ~=~ \int_{x=0}^a \int_{y=0}^a \psi^{(b)} (x, y) dx dy 
$$
There are somewhat different ways to define the fourth entry; for example,
$$
   k a \Psi^{DB}_{31} ~=~ \Deltarm_x \langle d_x \rangle - \Deltarm_y \langle d_y \rangle
$$
The actual calculation of the first column yields
$(0.5 \iu (k a)^{-1}, -0.5 \iu (k a)^{-1}, 0.25, 0)^T$,
the other three columns being very similar.

Solving for the $4 \times 4$ material parameter matrix $\eta$, one verifies that,
as expected, its only nonzero entries are $\epsilon_{xx} = \epsilon_{yy}$ and $\mu$.
The ``spatial dispersion block'' -- in this case, a single column \#4 of the matrix --
is zero, indicating the absence of ``spatial dispersion'' for a homogeneous cell,
as also expected. The effective parameters are not exactly equal to one but rather
$\epsilon_{xx} = \epsilon_{yy} \approx 1 + 5 \cdot 10^{-5} (k a)^{-1}$,
$\mu \approx 1 + 3.5 \cdot 10^{-5} (k a)^{-1}$.
The small deviation from unity is a result of the approximations involved -- namely,
of the free-space fields in the cell by spatially bilinear basis modes.

%
\subsection{Inclusions with Interior Resonances}\label{sec:Example-interior-resonances}
%
This example involves high-permittivity inclusions with strong resonance effects
and elements of spatial dispersion. The setup and parameters
are very similar to the ones in \cite{Felbacq-PRL05,Felbacq-MOTL09}, where a special asymptotic
method was developed to handle resonances and additional adjustable parameters
\cite{Felbacq-MOTL09} were needed. The methodology of the present paper handles this case
with relative ease and without any additional assumptions or parameters.

The lattice cell contains a high-permittivity  ($\epsilon_\mathrm{incl} = 200+5\iu$)
cylindrical inclusion; its radius relative to the cell size is $r / a = 0.25$;
the vacuum wavelength $\lambda / a$ varies from 5 to 12. It is interesting to
consider the $H$-mode ($p$-mode) where the electric field cuts across the cylinders
and strong resonances can be induced.

\begin{figure}
    \centering
   \includegraphics[width=0.85\linewidth]{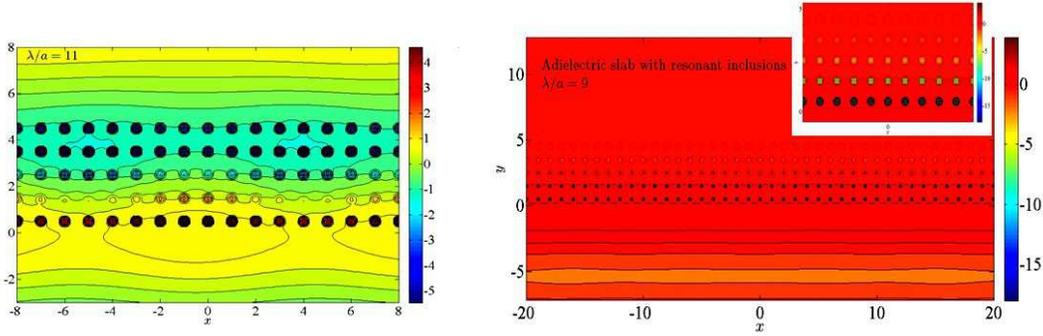}
\caption{{\small Re$(H)$ in the vicinity of a slab with resonant inclusions; $p$-mode. Radius of the inclusions
   relative to the cell size $r / a = 0.25$; their permittivity $\epsilon_\mathrm{incl} = 200+5\iu$
   (as in \cite{Felbacq-MOTL09}). Left: pass band, $\lambda / a = 11$; right: bandgap,
   $\lambda / a = 9$; inset: zoom-in on a few cells.}}
\label{fig:ReH-near-Felbacq-slab}
\end{figure}

\begin{figure}
\centering
    \includegraphics[width=0.45\linewidth]{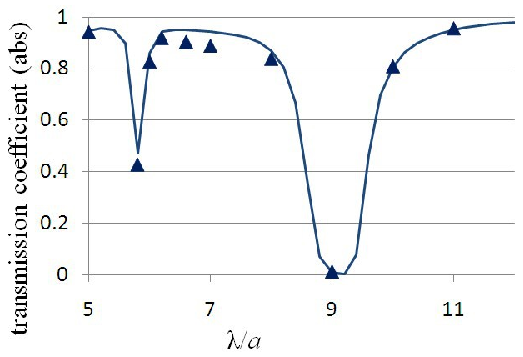}
    \includegraphics[width=0.45\linewidth]{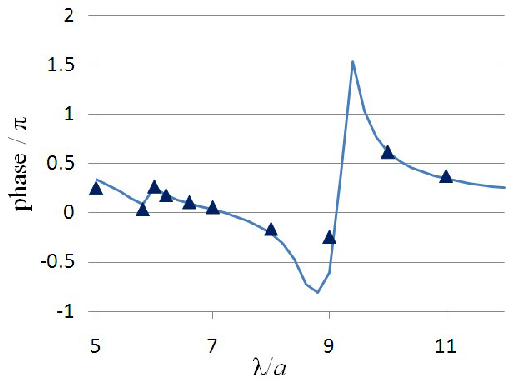}
   \caption{{\small Transmission coefficient (with respect to $H$) of a five-layer slab as a function
   of the vacuum wavelength (top: absolute value, bottom: phase).
   $r / a = 0.25$; $\epsilon_\mathrm{incl} = 200+5\iu$.
   Triangles: direct finite difference simulation \cite{Tsukerman-book07}; lines:
   slab with effective parameters. Normal incidence.}}
\label{fig:Felbacq-PhC-transmission}
\end{figure}

The effective medium theory developed here agrees very well with ``brute force''
finite difference simulations where all inclusions are represented directly.
As in \cite{Tsukerman-JOSAB11}, propagation of waves through a homogeneous ``effective parameter''
slab is compared with the numerical simulation of this propagation through the actual metamaterial;
high-order finite difference ``FLAME'' schemes (\cite{Tsukerman06,Tsukerman-book07,Tsukerman-PBG08}
were used for the numerical simulation. For illustration, \figref{fig:ReH-near-Felbacq-slab}
shows the color plots of the real part of the magnetic field for normal incidence
in the pass band, $\lambda / a = 11$, and in the bandgap,  $\lambda / a = 9$.

\figref{fig:Felbacq-PhC-transmission} demonstrates that the transmission coefficient
for the slab with effective parameters is very close, both in the absolute value and phase,
to the ``true'' coefficient from the accurate finite difference simulation.

\begin{figure}
\centering
   \includegraphics[width=0.4\linewidth]{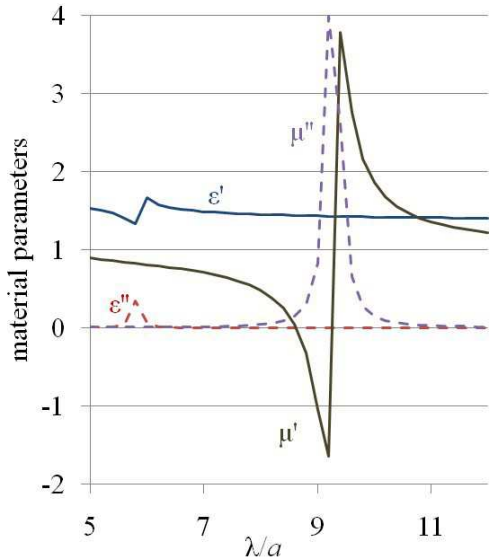}
\caption{{\small
   Effective parameters of the metamaterial with resonant inclusions at the center of the lattice cell.}}
\label{fig:effective-params-Felbacq-cyl-PhC}
\end{figure}

The effective parameters plotted in \figref{fig:effective-params-Felbacq-cyl-PhC}
are completely physical; they do exhibit physical resonances but no ``antiresonances''.
A fairly weak electric resonance is observed near $\lambda / a \sim 5.5$
and a strong magnetic resonance  -- around $\lambda / a \sim 9$.
As should be expected by symmetry considerations, the magnetoelectric coupling
is in this example zero (numerically, it is at the roundoff error level).

\begin{figure}
\centering
  \includegraphics[width=0.6\linewidth]{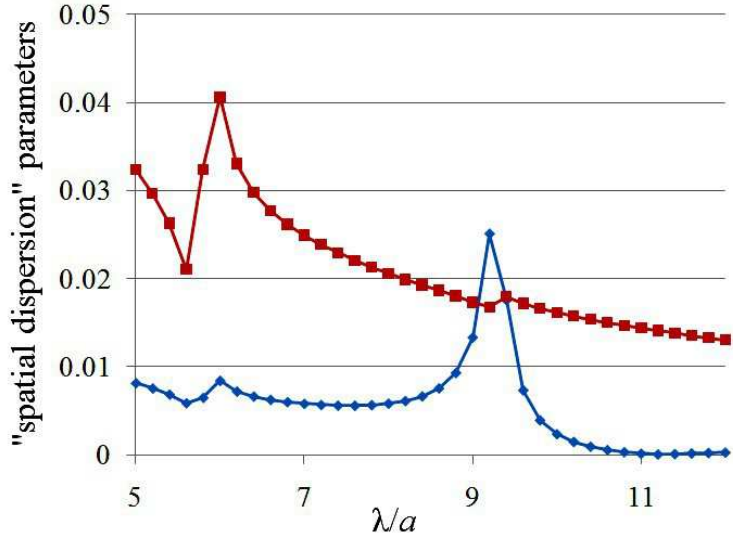}
\caption{{\small
   The absolute values of the ``spatial dispersion'' parameters $\eta_{14} = \eta_{24}$ (squares)
   and $\eta_{34}$ of the metamaterial with resonant inclusions.
   Shifted inclusions centered at $x_0 = y_0 = 0.1a$ relative to the cell center.}}
\label{fig:spatial_dispersion_params_Felbacq_shifted_cyl.eps}
\end{figure}

The entries $\eta_{14}$, $\eta_{24}$ and $\eta_{34}$ of the parameter matrix
-- measures of nonlocality -- are in this case zero, too. However, this is no longer so if the inclusions
are shifted to, for instance, $(x_0, y_0) = (0.1a, 0.1a)$
relative to the center of the cell. Parameters $\epseff$ and $\mueff$ are almost
unaffected by this shift (third- or fourth-digit differences). However,
$\eta_{14} = \eta_{24}$ and $\eta_{34}$ now have absolute values
on the order of 0.05 or less (\figref{fig:spatial_dispersion_params_Felbacq_shifted_cyl.eps}),
and going down to zero in the long-wavelength limit, as expected.

This weakly nonlocal response is consistent with physical intuition and is clearly
due to the appreciable size of the cell relative to the vacuum wavelength
and especially to the wavelength in the inclusions.
Why, though, does not the nonlocality manifest itself when the inclusions
are located at the center of the cell?
Nonlocality reflects the fact that the behavior of the fields within the cell
is too complex to be described (at a given level of approximation)
just by the field averages; more information is needed and
comes in the form of field variations.
However, when the cell is symmetric, fewer parameters may suffice to describe
the field (this is akin to, say, the permittivity tensor degenerating to
a single scalar value in the symmetric case); hence one can make do without any
additional nonlocality parameters unless a much higher level of approximation is desired.

%
%
%
\section{Conclusion}\label{sec:Conclusion}
%
The proposed homogenization theory stems from a small number of fundamental principles --
most importantly, that the coarse-grained fields must satisfy Maxwell's equations and
boundary conditions and that material parameters are, by definition,
linear relations between the coarse-grained fields. Consequently, the $\bfE$ and $\bfH$ fields are
produced by an interpolation that preserves tangential continuity
(see \sectref{sec:Proposed-theory-details} and \cite{Tsukerman-JOSAB11}), while for
$\bfD$ and $\bfB$ the normal continuity is maintained.

The number of degrees of freedom (d.o.f.) for the coarse-grained fields depends on the
specific interpolation procedures chosen. (For example, the most natural choice for a parallelepipedal
lattice cell results in 24 d.o.f. for the $(\bfE, \bfH)$ pair and 12 d.o.f. for the $(\bfD, \bfB)$ pair.)
Technically, therefore, the generalized ``material parameter'' can be represented by a large matrix
that encodes a substantial amount of information about the behavior of the fields in the cell.
This parameter matrix acquires a clear physical meaning upon transformation to
a ``canonical'' basis where the coarse-grained fields are expressed as their average
values plus variations. In this canonical representation, the matrix
has a leading $6 \times 6$ diagonal block that represents the standard effective
parameters: the permittivity, permeability and magnetoelectric tensors.
The remaining matrix blocks -- linking the average values of the coarse-grained fields
to their variations over the cell -- may serve as a quantitative measure of spatial dispersion.

The new theory is a substantial extension of the methodology put forward
in \cite{Tsukerman-JOSAB11}. The overall structure of the procedure
(cf. \figref{fig:Scheme-proposed-methodology}) and its core -- two different types
of interpolating functions for different fields -- remain the same.
However, the linear relations between the coarse-grained fields are now established
and handled differently, allowing one to evaluate not only the traditional electromagnetic
parameters but also some quantitative measures of spatial dispersion as described above.
Furthermore, this paper breaks away from the exclusive reliance on Bloch waves
as approximating modes for the field. While propagating Bloch modes are indispensable in
the analysis of periodic structures, their value for nonperiodic media and
especially for evanescent waves is limited; in addition, Bloch modes are expensive to compute.
One possible alternative is low-order polynomial approximation of the fields on lattice cell boundaries.

The proposed theory does not involve any heuristic assumptions or artificial averaging rules.
The effective parameters are defined directly via field analysis in the lattice cell,
in contrast with methods where these parameters are obtained
from reflection/transmission data or other indirect considerations.
Nontrivial magnetic behavior, if present, follows logically from the method.
The necessary approximations are clearly identifiable and measurable.
Examples include a tutorial on the method and an analysis of a resonant structure
with high-permittivity inclusions.

%
\section*{Acknowledgment}\label{sec:Acknowledgment}
%
I thank Vadim Markel, Dmitry Golovaty, Graeme Milton, Boris Shoykhet and Sergey Bozhevolnyi
for very helpful and insightful comments and discussions. Many thanks to Anders Pors
for implementing the method in 3D \cite{Pors11} and for asking pointed questions
that helped to crystallize the ideas of the present paper.

\bibliographystyle{plain}


\end{document}